# Quadratic-Gradient Metasurface-Dome for Wide-Angle Beam Steering Phased Array with Reduced Gain-Loss at Broadside

Alessio Monti, *Senior Member, IEEE*, Stefano Vellucci, *Member, IEEE,*
Mirko Barbuto, *Senior Member, IEEE,* Davide Ramaccia, *Senior Member, IEEE,* Michela Longhi,
Claudio Massagrande, Alessandro Toscano, *Senior Member, IEEE,* Filiberto Bilotti, *Fellow, IEEE*

*Abstract*—Recently, the use of deflecting covering meta-domes for increasing the scanning range of a phased array has been proposed. Although promising, this solution still presents some inherent limitations such as the significant reduction of the broadside gain compared to the original array, as well as the complexity of the implementation due to the need for a continuous phase profile. This work aims at proposing some technical solutions for maximizing the meta-dome performance and reducing the gain-loss at broadside. By properly discretizing and engineering the phase profile along the dome and by taking into account the different angles of incidence onto the meta-cells, we show how it is possible designing realistic meta-domes with reduced insertion loss at broadside, improved steering capabilities, and a reduced profile.

## I. INTRODUCTION

Beam-scanning antennas play a crucial role in many applications such as radar, satellite, and point-to-point communication and are expected to be even more important for beyond 5G and 6G wireless systems [1]-[3]. Among the different technical solutions used to achieve beam-scanning, phased antenna arrays still represent the most consolidated and adopted technique. Unfortunately, the scanning range of a phased array – defined as the maximum pointing angle for which the array gain exceeds a desired threshold - is limited by several factors. For example, the half-power beam-width (HPBW) of conventional printed antennas that is usually limited to 60° [4], and the mutual coupling between consecutive elements, which increases progressively with the scan angle [5].

To overcome these issues, several techniques have been discussed in the last years based, for instance, on the use of wide-beam [6]-[7] or reconfigurable antenna elements [8]-[9]. Still, in these cases, a re-design of the radiating element is required. Wide-angle impedance matching (WAIM) aiming at reducing the mutual coupling between the array elements for large scanning angles have been also proposed [10]-[12]. Here, the impedance matching of the radiating elements during the array scanning is improved, but have a moderate effect on the array directivity and gain.

More recently, a new approach based on the deflection of the beam emitted by the array for a given pointing angle through a local gradient of the transmission phase has been discussed [13],[14]. This approach requires the use of an external dome, either made of a phase-gradient surface (such as the ones described in [15]-[22]) or by a properly shaped dielectric, and can be applied to any pre-existing system. Remarkably, the meta-dome approach allows increasing the scan range for low-elevation angles without introducing additional grating lobes in the visible spectrum. However, here, the increase of the scan range is obtained at the expense of a significant reduction of the broadside gain, because any surface extending the scan-range also behaves as a diverging lens in the broadside direction [23]. Moreover, a continuous phase variation is required, which is not easy to be implemented in practical designs.

In this work, we propose the design of a realistic phase-gradient metasurface-dome with reduced gain loss at broadside and improved deflecting performance. Through a careful analysis of the array near-field distribution and of the discretization strategy of the phase insertion, we show that a quadratic profile allows significantly reducing the broadside insertion loss of the dome and increasing its steering capabilities.

Some very preliminary results in this direction have been discussed in our recent conference paper [24]; the main novel elements of this manuscript can be summarized as follows:

a) Phase discretization on the performance of the dome is investigated, and it is shown that a quadratic-gradient allows reducing the broadside insertion loss of the dome whilst improving the overall scan range of the system;

b) A complete design workflow for realistic meta-domes composed of Huygens meta-cells working under oblique incidence and for any polarization of the original array is proposed and verified through full-wave simulations.

Manuscript received February 01, 2022. Revised July 09, 2022. Revised September 08, 2022. Accepted November 07, 2022.

This work has been developed in the frame of the activities of the Project MANTLES, funded by the Italian Ministry of University and Research under the PRIN 2017 Program (protocol number 2017BHFZKH). (Corresponding author: Alessio Monti, alessio.monti@uniroma3.it).

A. Monti, S. Vellucci, D. Ramaccia, A. Toscano, and F. Bilotti are with the Department of Industrial, Electronic, and Mechanical Engineering, ROMA TRE University, 00146 Rome, Italy.

M. Barbuto, and M. Longhi are with the Department of Engineering, Niccolò Cusano University, 00166, Rome, Italy.

C. Massagrande is with Milan Research Center - Huawei Technologies, Milan, Italy.

## II. Design Fundamentals

### A. Phase insertion discretization

Fig. 1(a) shows the reference geometry considered in this paper. It consists of a linear phased array of eight printed antennas with an inter-element separation distance equal to $0.47\lambda_0$. The antennas used in this structure are conventional patches exhibiting a slanted $\pm45°$ dual-linear polarization and a half-power beamwidth on the horizontal ($xz$-) plane of 80°.

The array is covered by a dome with maximum height equal to $h$. The dome consists of discrete Huygens meta-cells able to introduce a different phase-gradient to the beam radiated by the array. The aim of the dome is to improve the grating-lobe free scan range of the array by steering toward the horizon the beam radiated by the array for its maximum pointing angle $\theta_{in}$.

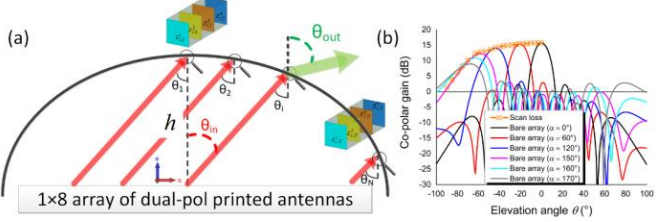

Fig. 1: (a) Reference geometry considered in this paper consisting of a phase-gradient meta-dome applied to a linear array of dual-polarized printed antennas. Inset: met-atoms composing the dome consisting of four surface impedances separated by three identical dielectric slabs. (b) Co-polar gain on the horizontal ($xz$) plane of the bare array for different values of the inter-element phase shift α. The orange line represents the array scan loss. The horizontal line represents the 0 $dB$ threshold.

To better explain this point, in Fig. 1(b) we report the radiation diagram on the horizontal plane of the bare array, *i.e.*, without (w/o) the dome, for different values of the excitation phase shift between radiating elements $\alpha$ and for the -45° pol. These results have been obtained through full-wave simulations and considering periodic boundary conditions along the *y*-axis. As can be inferred, because of the reduced beamwidth of the printed antennas, the maximum co-polar gain of the array reduces as the pointing angle is progressively increased and scan-loss appears. Here, the maximum pointing angle of the array $\theta_{in}^{max}$ is defined by setting a threshold of 0 dB for the grating lobes, which are known to increase as the phase shift between antenna elements increases. In this scenario, $\theta_{in}^{max} = -58°$ and is achieved for a phase shift between the array elements $\alpha_{max}= 150°$.

The scanning performance of the array can be improved using a deflecting meta-dome placed on the top of the array, as shown in Fig. 1(a). According to [13], the required phase insertion along the meta-dome profile can be calculated using the local version of the generalized law of refraction [25] and assumes the following expression:

$$\Phi(x) = \Phi_0 + k\int_0^x f(\xi)\sqrt{1+\left(g'(\xi)\right)^2}\,d\xi, \qquad (1)$$
$$f(x) = \sin\left[\gamma_i\left(\vartheta_{out}\right)\right] - \sin\left[\gamma_o\left(\theta_{in}^{max}\right)\right],$$

where *x* is the geometrical coordinate, *k* is the wavenumber, $z =$ $g(x)$ is the function describing the geometrical shape of the meta-dome, $\theta_{out}$ is the desired output angle from the meta-dome, $\Phi_0$ is an arbitrary phase value assumed as reference, and $\gamma_{i,o}$ are the angles that the input and output wavevectors form with the normal direction. Please note that phase insertion $\Phi(x)$ is an even function and, therefore, the dome introduces the same beam steering for positive and negative values of $\theta_{in}$. Indeed, the phase insertion $\Phi(x)$ is a continuous function that needs to be discretized in a finite number 2N of samples for practical implementation.

To better understand the effect of the phase discretization, let's consider a simple semi-circular meta-dome where the geometrical profile of the dome can be thus written as $g(x) = -z_0 + \sqrt{R^2 - x^2}$, being *R* the radius of the meta-dome. We assume that the center of the meta-dome coincides with the center of the array, *i.e.*, $z_0 = 0$, that the radius of the meta-dome is $R = 2\lambda_0$, and thus $h = 2\lambda_0$. Moreover, we set the input and output angles such that the desired additional steering introduced by the meta-dome is equal to $\Delta\theta = \theta_{out} - \theta_{in}^{max} = 15°$.

In Fig. 2(a), we show the phase insertion function of this configuration, calculated through (1), and compare two different sampling strategies assuming a number of samples equal to $N = 12$. The first one corresponds to a standard uniform gradient, *i.e.*, each pair of consecutive samples $x_i$ and $x_{i+1}$ exhibits the same phase difference $\Phi(x_{i+1}) = \Phi(x_i) = const$. The second sampling strategy, instead, satisfies a quadratic profile. The geometrical position of each sample for these two approaches can be calculated, respectively, as:

$$x_i^{linear} = \pm\frac{L}{2}\frac{i}{(N-1)}, \quad i = 0,1,...N-1,$$
$$x_i^{quadratic} = \pm\frac{L}{2}\frac{i^2}{(N-1)^2}, \quad i = 0,1,...N-1. \qquad (2)$$

being *L* the semi-length of the meta-dome.

As can be observed in Fig. 2(a), the quadratic profile allows reducing the phase difference between consecutive samples on the top of the meta-dome and magnifying it on the lateral parts of the structure. We stress that the width of the cells is the same in both the scenarios: therefore, a quadratic sampling strategy effectively result in a quadratic phase profile along the dome.

The two domes exhibiting the phase profile shown in Fig. 2(a) have been implemented through realistic unit-cells and simulated with a full-wave software. The details of the practical implementations of the *N* Huygens cells composing the meta-dome are described in the next sections. The co-polar gain of the array on the horizontal plane for different values of *α* are reported in Fig. 2 (b). The dashed lines refer to the scenario in which the array is covered by the designed meta-dome, whereas the continuous lines refer to the results obtained for the original (bare) array.

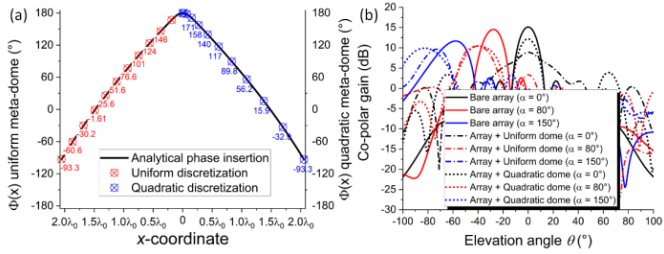

Fig. 2: (a) Phase insertion $\Phi(x)$ of the designed dome (continuous black line) and comparison between uniform (red ticks) and quadratic (blue ticks) discretization. For each configuration, the samples for negative values of $x$ can be easily calculated by symmetry, i.e., $\Phi(x_i) = \Phi(-x_i)$. (b) Co-polar gain on the horizontal plane of the bare array (continuous lines), of the array covered by a dome exhibiting a uniform-discretized phase profile (dashed lines) and for a quadratic-discretized phase profile (dotted lines). The different lines refer to different values of the inter-element phase shift $\alpha$.

From these results, it is evident that the quadratic phase profile allows reducing significantly the insertion loss (IL) of the meta-dome when the array points broadside ($\alpha=0°$), where the IL is defined as the difference of the co-polar gain of the array with and without the dome in broadside direction. In particular, for this case study, the IL is reduced from 6 dB (uniform-gradient meta-dome) to 3 dB (quadratic-gradient meta-dome).

The get a better understanding of the advantages of the quadratic phase profile over the uniform, it is interesting to observe the distribution of the electric field onto the semi-circular meta-dome. As reported in Fig. 3 (a), when the array points broadside ($\theta_{in} = 0°$), the energy radiated by the array is spread throughout the entire meta-dome but it is mainly concentrated on its central region. Since the quadratic profile exhibits a *reduced phase difference* between consecutive samples in the central region of the dome, the IL and the diverging effect is significantly reduced compared to the uniform case.

Remarkably, also the steering performances of the quadratic meta-dome are improved compared to the one obtained in the uniform scenario. In particular, as shown in Fig. 2(b), when the array is excited to point at $\theta_{in}^{max}$ (i.e., for $\alpha_{max}=150°$), an additional tilting of 25° is introduced by the dome (i.e., $\vartheta_{out} = -83°$ vs. $\theta_{in}^{max} = -58°$). Conversely, the additional tilting introduced by the uniform-gradient meta-dome is lower and limited to 15° ($\vartheta_{out} = -73°$ vs. $\theta_{in}^{max} = -58°$). Also in this case, the result can be easily understood by considering the field distribution in Fig. 3(a). When the beam radiated by the array is steered towards the horizon ($\theta_{in} > 0°$), the energy is progressively focused on the lateral regions of the dome. In these regions, the quadratic meta-dome exhibits a *higher phase gradient* compared to the uniform meta-dome and, as such, it introduces a higher deflecting effect on the impinging beam. From this discussion, it should be clear that, when dealing with a quadratic-gradient meta-dome, its steering angle is higher than the design value in (1). Therefore, the additional steering angle used to compute $\Phi(x)$ should be lower than the desired one to compensate for the increased steering performance of the quadratic meta-dome.

We remark here that, by changing the exponent in (5), the phase profile can be finely controlled and the IL at the broadside could be further reduced. However, this is generally paid by a worsening of the steering performance for intermediate pointing angles. Therefore, we can conclude that the quadratic-gradient discretization returns a reasonable trade-off between the unavoidable perturbation introduced by meta-dome to the array broadside radiation and its overall steering performance.

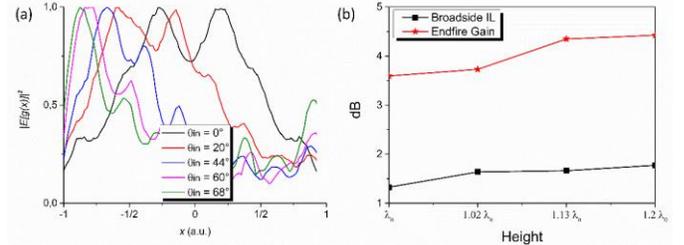

Fig. 3: (a) Normalized electric-field distribution along the meta-dome profile for different angles of illumination. (b) Broadside IL and end-fire gain as a function of meta-dome height.

### B. Effect of the height on the dome performance

Although a meta-dome with a quadratic phase profile has a lower IL compared to a uniform-one, its height $h$ affects significantly its performance. As a preliminary consideration, we observe that, once fixed the maximum pointing angle of the array $\theta_{in}$, there exists a minimum value for the height dome $h$ that ensures its proper working. According with [14],[26], this condition can be expressed as $h_{min} = \frac{L}{2} \cot \vartheta_{in}^{max}$, being $L$ the length of the array.

In our scenario, the phased array consists of eight printed antennas with an inter-element separation distance equal to $0.47\lambda_0$ and with $\theta_{in}^{max} = -58°$, which returns, according to the previous equation, $h_{min} \approx \lambda_0$. Indeed, it is interesting what is the effect of a thickness $h > h_{min}$. For this purpose, we have compared the performances of four semi-circular meta-domes characterized by a height $h$ contained in the range [$\lambda_0$ - 1.2 $\lambda_0$], and introducing an additional beam steering of $\Delta\theta = 10°$ for $N = 12$ discrete samples.

Fig. 3 (b) shows the broadside IL and the end-fire gain as a function of the height of the meta-dome. The minimum IL is obtained for the lowest height value ($h = \lambda_0$). This configuration also led to the lower value of end-fire gain. The opposite behaviour is obtained for a larger value of the height: in the highest meta-dome scenario, the broadside IL is maximized, whilst the end-fire gain is enhanced. These results can be easily understood by considering again the near-field distribution shown in Fig. 3(a) and the phase insertion function of the meta-dome shown in Fig. 2(a). Because of the shape of the dome, when the array points at the broadside, a wider region of the dome is illuminated as its height is progressively increased. Since the phase function is symmetric with respect to the dome axis, its undesired diverging effect increases with the height. On the contrary, because of the quadratic phase profile, an increase of the dome height also increases the phase gradient experienced by the beams radiated by the array for large pointing angles. Therefore, the directivity in end-fire direction increases with the dome height.

This analysis confirms that the choice of the dome height holds particular importance, even when a quadratic phase profile is considered. Depending on the specific applications, it is possible to choose either a low-profile dome to minimize broadside IL as much as possible or thicker structures if the primary scope is to improve the directivity for extreme elevation angles.

*C. Angle of incidences*

The use of a meta-dome with a curved profile necessarily implies that the cells implementing the dome are excited in off-normal conditions. Therefore, to implement the required phase profile, we need to account for the different angles of incidence of the beam emitted by the array onto the discrete cells.

The reference scenario is shown in Fig. 1(a). Assuming a plane-wave approximation for the beam radiated by the array, from simple geometrical considerations the incidence angles $\theta_i$ onto the cells (with $i = 0, 1,…, N-1$) can be calculated as:

$$\theta_i = \theta_{in}^{max} - \arctan\left[\frac{\partial}{\partial x} g(x)\right]_{x=x_i} + 90°, \quad (3)$$

where $x_i$ is the coordinate of the center of each discrete cell (which depends on the chosen discretization profile) and $g(x)$ is the analytical function describing the dome shape.

Once the incidence angles are known, we can design each discrete cell able to exhibit the required phase insertion $\Phi(x_i)$. The meta-atom considered here is shown in the inset of Fig. 1(a). It consists of four impedance sheets ($Z_{s,i}^1, Z_{s,i}^2, Z_{s,i}^3, Z_{s,i}^4$) separated by three identical dielectric layers of thickness $d$ and relative permittivity $\varepsilon_r$. This geometry ensures the co-excitation of electric and magnetic dipoles and, therefore, allows achieving high transmission values and wide phase coverages [27]-[29]. Although just three impedance sheets can be used to design Huygens metasurfaces matching the different wave impedances at the sides of the cells [19], a fourth layer is introduced as an additional degree of freedom.

From a modelling point of view, the responses of the individual cells can be calculated using conventional microwave matrices [30]. In our scenario, however, there is an additional complexity resulting from the off-normal excitation and from the dual-polarization operation of the array that requires each cell to have the same behaviour for a specific angle of incidence. A numerical optimization procedure, aiming at finding the optimum combination of surface impedances for each of the $N$ discrete cells able to guarantee the desired transmission phase along the dome, can be implemented starting from the analytical expressions of the transmission coefficients for the two polarizations. Details are here omitted for the sake of brevity and can be found in [31].

## III. Design of a Beam-Steering Meta-dome with Reduced Height

In the previous Sections, we have discussed several design strategies for improving the overall performance of a deflecting metasurface-dome aimed at increasing the scanning range of a planar array without compromising the gain value at broadside. To validate our insights, we consider here a case of practical interest, where all the discussed design fundamentals have been applied for maximizing the system performance.

At first, an ideal meta-dome composed by unit-cells exhibiting the analytically designed values of surface reactance is presented. Then, a realistic structure considering the practical realization of the cells is reported.

*A. Ideal meta-dome*

With the aim of minimizing the degrading effect of the meta-dome on the broadside radiation of the array and, at the same time, improving its steering capability, an ideal meta-dome with a deflecting angle of 10°, $R = 2.5\ \lambda_0$, $z_0 = 1.4\ \lambda_0$, and $h = \lambda_0$ has been considered. The unit-cells of the meta-dome have been designed for implementing a quadratic phase profile and by considering the different angles of incidence depending on the scanning angle of the array, as well as the presence of a radome for structural stability and protection from environmental effects has been introduced.

A sketch of the structure is shown in Fig. 4(a). The Huygens cells consists of three foam spacers ($\varepsilon_r = 1.2$, $tan\delta = 0.001$) with a thickness $d = \lambda_0/29$, whilst the radome substrate on top is made of a common commercially available material ($\varepsilon_r = 2.55$, $tan\delta = 0.004$) of thickness $t = \lambda_0/35$. The thickness of the dielectric has been chosen to maximize the phase coverage returned by the different combinations of $Z_{s,i}^1, Z_{s,i}^2, Z_{s,i}^3, Z_{s,i}^4$.

The final performances of the selected cells are summarized in Table I. In the second column we report the values of the impinging angles onto each cell when the array steers towards 58°. The desired phase shift of the individual cell - derived from (1) with the quadratic phase profile reported in (2) - are reported in the last column, whereas the actual values obtained optimizing the cells with the semi-analytical procedure outlined above are reported in columns #4-#7. It is worth mentioning

TABLE I

Angle of incidence and transmission coefficients of the 12 ideal cells when

| Cell# | $\theta_i$ | $\|S_{21}^{i\ TE}\|$ | $\|S_{21}^{i\ TM}\|$ | $\angle S_{21}^{i\ TE}$ | $\angle S_{21}^{i\ TM}$ | $\Phi(x_i)$ |
|---|---|---|---|---|---|---|
| 0 | 148° | 0.84 | 0.85 | 6° | -6° | 0° |
| 1 | 144° | 0.86 | 0.85 | 7° | -5° | -0.1° |
| 2 | 139° | 0.93 | 0.85 | 0° | -1° | -1.8° |
| 3 | 135° | 0.95 | 0.90 | -5° | -7° | -5° |
| 4 | 131° | 0.95 | 0.92 | -12° | -13° | -12° |
| 5 | 126° | 0.92 | 0.91 | -22° | -22° | -21° |
| 6 | 122° | 0.91 | 0.92 | -30° | -29° | -32° |
| 7 | 118° | 0.92 | 0.94 | -37° | -36° | -46° |
| 8 | 114° | 0.92 | 0.94 | -61° | -58° | -63° |
| 9 | 110° | 0.99 | 0.99 | -80° | -76° | -84° |
| 10 | 106° | 0.97 | 0.97 | -103° | -98° | -108° |
| 11 | 101° | 0.84 | 0.83 | -134° | -132° | -138° |

that the surface reactance values $\chi_{s,i}^j$ of the different layers of the 12 cells implementing the desired phase shift are in the range +300/-300 Ω/sq., ensuring feasibility of the realistic design.

The effects of the dome on the scanning performance of the array are shown in Fig. 4 (b), where the co-polar gain on the horizontal plane of the array for the cases of bare and covered array are shown. We can observe that the deflecting properties of the meta-dome allow increasing the maximum steering angle by 10°, and the end-fire gain for α=150° by 4.6 *dB*. At the same time, the broadside IL introduced by the meta-dome was limited to around 1.5 *dB* as a result of the selected quadratic discretization. Please note that, by optimizing the excitation coefficients of the array elements in both amplitude and phase [13], further improvements of the overall performance can be obtained and, in particular, the IL at broadside can be further reduced by almost *0.5 dB*.

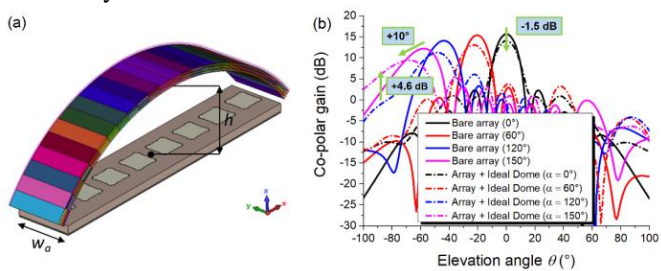

Fig. 4: (a) Geometrical sketch of the proposed meta-dome covering a standard dual-polarized patch antenna array. (b) Co-polar gain on the horizontal plane of the bare array (continuous lines) and of the array covered by the meta-dome implemented trough ideal Huygens cells (dashed lines). The broadside IL is around 1.5 *dB*, while the endfire gain has improved by *4.6 dB*.

### B. Real meta-dome

For the physical implementation of the ideal Huygens cells, we have exploited a two-step optimization procedure. First, the surface impedance values obtained in Section III-A have been used to design each individual layer, according with the available homogenization formulas and retrieval procedure (see, for instance, [32]). However, when the four layers are cascaded to build a Huygens meta-atom, the coupling between them modifies their individual response. Therefore, a full-wave optimization procedure is required to finitely re-optimize their overall geometry.

The stack-up of the square cells considered is reported in Fig. 5. As for the ideal meta-cells design, the realistic stack-up is made of three foam layers ($d = \lambda_0/29$, $\varepsilon_r = 1.2$) and a radome substrate on top ($t = \lambda_0/35$, $\varepsilon_r = 2.55$), whilst the length of the square shaped cell is $w_a/4$. The layout of the metal patterns has been chosen for satisfying the dual-pol requirement (*i.e.*, returning an isotropic layout), and are made of a square patch with an external square ring [27]. The geometry on the four layers have been optimized through a full-wave numerical procedure with the aim of reproducing the complex-valued transmission coefficient values reported in Table I (column 3-6). The final values of $d_c^{1,2,3,4}$ and $w_c^{1,2,3,4}$ are in the range of $[\lambda_0/226 - \lambda_0/5.43]$ and $[\lambda_0/770 - \lambda_0/250]$, respectively.

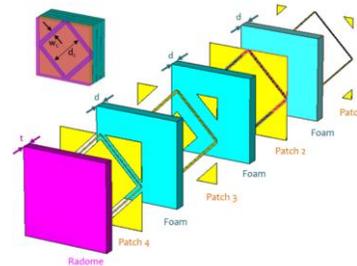

Fig. 5: Stack-up and its exploded vision of the realistic unit-cell consisting of three foam-like dielectric spacers plus the radome layer, and four square-shaped metallically patterned layers. In the Floquet-mode analysis, the cell is 45° tilted in order to replicate its effective geometry when placed in the meta-dome.

It is observed that, in general, the magnitude of the transmission coefficient of all the real cells is lower than the one of the ideal cells (average value around 0.7). This is mostly due to two limiting points: the reduced number of samples exploited in the optimization procedure *w.r.t.* the one considered in the analytical design; the minimum dimensions of $d_c$ and $w_c$ that have been settled to be compliant with fabrication constrains.

The final structure of the quadratic-gradient meta-dome implemented through the realistic cells is reported in Fig. 6(a)-(b). As shown in the inset, to fill the empty gaps on the higher layers of the cells due to the different experienced curvature, thin metallic strips between adjacent cells have been introduced to allow for continuity of the surface currents and minimum deterioration of the meta-dome performance compared to the ideal scenario. It is worth mentioning, however, that in this way the single cell response is different compared to the one from numerical optimization, due to the slight variation in the dimension of the external metallic rings, especially for the top layer. However, this discrepancy is expected to be quite small due to the extremely small dimension of the added strips.

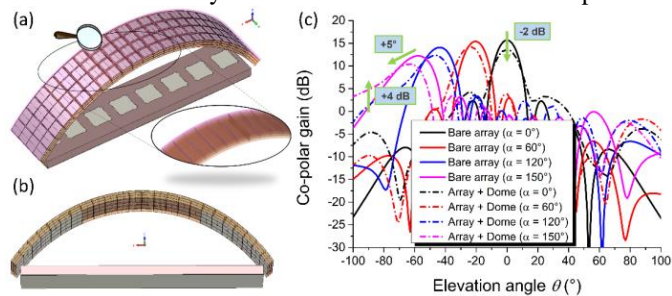

Fig. 6: (a)-(b) Sketches of the proposed meta-dome implemented through realistic unit-cells. (a) Perspective and (b) bottom views (in (a) the radome layer is transparent). In the inset, details of the metallic strips filling the gaps between cells. (c) Co-polar gain on the horizontal plane of the bare array (continuous lines) and of the array covered by the meta-dome implemented trough realistic Huygens cells (dashed lines). The broadside IL is around 2 *dB*, while the end-fire gain is improved by *4 dB*. As reported for the ideal-cell scenario, the overall performance can be further improved by an average of *0.5 dB*, by properly optimizing the excitation coefficients of the array elements [14].

The final performance of the designed meta-dome is reported in Fig. 6(c), where the co-polar gains on the horizontal plane of the array for the cases of bare and covered array are shown. It can be observed that the end-fire gain improvement introduced by the dome for α=150° is equal to 4 *dB*. At the same time, the deflecting properties of the meta-dome allow increasing the maximum steering angle by +5°, with an insertion loss of *2 dB* when pointing at the broadside.

Compared to the performance of the ideal meta-dome configuration in Fig. 4, we can observe a slight deterioration. Remarkably, the co-polar gain difference between the ideal-cells meta-dome and the realistic configuration is quite limited (*0.5 dB* at the broadside, and *0.6 dB* when steering towards 90°). Indeed, the slight deterioration of the performance compared to the ideal structure is due to the worse performance of the real cells compared to the ideal ones in terms of both amplitude of the transmission coefficients and phase shifts achieved compared to the desired ones, and the transformation of the planar cells into curved ones.

## IV. Conclusions

We have discussed several advanced design strategies for metasurface-based dome aimed at enhancing the beam steering capability of dual-pol phased arrays. Compared to earlier works, the metadevices proposed here are characterized by low gain loss at the broadside, improved steering performances, minimum height, and a fabrication-ready design. Indeed, the meta-domes discussed in this paper have been fully implemented through curved multi-layered metasurfaces made of Huygens cells able to return the phase transmission needed to deflect the beam radiated by the array towards the horizon regardless its polarization.

These results validate and further emphasize, the potentialities of meta-dome devices to improve the scanning capabilities of phased arrays, reducing the scan loss for extreme angles, without severely compromising the performance at the broadside. The proposed device can be particularly useful for 5G and beyond 5G applications, in view of array systems characterized by wide-angle beam steering able to overcome the scanning limitations of conventional phased arrays.